\documentclass[sigconf, 10pt,authorversion,nonacm]{acmart}
\usepackage{graphicx} 
\PassOptionsToPackage{prologue,table}{xcolor}
\usepackage{comment}
\usepackage{soul}
\usepackage{xcolor}
\usepackage{ulem}
\usepackage{graphicx}
\usepackage{caption}
\usepackage{subcaption}
\usepackage{amsfonts}
\usepackage{amsmath}
\usepackage{mathtools, nccmath}
\usepackage{enumitem}
\usepackage[table]{xcolor}
\usepackage{color, colortbl}
\usepackage{multirow}
\usepackage{nicefrac}
\usepackage{amsbsy}
\usepackage{float}

\title{\textit{Real-Time Diagnostic Integrity Meets Efficiency}:\\ A Novel Platform-Agnostic Architecture for Physiological Signal Compression}

\author{Neel R Vora$^{\star,1,3}$, Amir Hajighasemi$^{\star,1}$, Cody T. Reynolds$^{1}$,
Amirmohammad Radmehr$^{3}$, Mohamed Mohamed$^{1}$, Jillur Rahman Saurav$^{1}$, 
Abdul Aziz$^{3}$, Jai Prakash Veerla$^{1}$, Mohammad S Nasr$^{1}$, 
Hayden Lotspeich$^{1}$, Partha Sai Guttikonda$^{1}$, Thuong Pham$^{1}$, 
Aarti Darji$^{1}$, Parisa Boodaghi Malidarreh$^{1}$, Helen H Shang$^{1,2}$, 
Jay Harvey$^{4}$, Kan Ding$^{4}$, Phuc Nguyen$^{\dagger,1,3}$, Jacob M Luber$^{\dagger,1}$}

\affiliation{%
  \institution{%
    $^{1}$ The University of Texas at Arlington, USA \\
    $^{2}$ Ronald Reagan University of California Los Angeles Medical Center, USA \\
    $^{3}$ The University of Massachusetts Amherst, USA \\
    $^{4}$ University of Texas Southwestern Medical Center, Department of Neurology
  }
  \country{USA}
}

\thanks{* These authors contributed equally to this work.}
\thanks{Corresponding author: Jacob M Luber (Email: jacob.luber@uta.edu) and Phuc Nguyen (Email: phuc@cics.umass.edu)}

\date{August 2023}

\begin{abstract}
    Head-based signals such as EEG, EMG, EOG, and ECG collected by wearable systems will play a pivotal role in clinical diagnosis, monitoring, and treatment of important brain disorder diseases. 
    However, the real-time transmission of the significant corpus physiological signals over extended periods consumes substantial power and time, limiting the viability of battery-dependent physiological monitoring wearables. 
    This paper presents a novel deep-learning framework employing a variational autoencoder (VAE) for physiological signal compression to reduce wearables' computational complexity and energy consumption. 
    Our approach achieves an impressive compression ratio of 1:293 specifically for spectrogram data, surpassing state-of-the-art compression techniques such as JPEG2000,  H.264, Direct Cosine Transform (DCT), and Huffman Encoding, which do not excel in handling physiological signals.  
    We validate the efficacy of the compressed algorithms using collected physiological signals from real patients in the Hospital and deploy the solution on commonly used embedded AI chips (i.e., ARM Cortex V8 and Jetson Nano). The proposed framework achieves a 91{\%} seizure detection accuracy using XGBoost, confirming the approach's reliability, practicality, and scalability. 
\end{abstract}

\begin{document}

\maketitle

\section{Introduction}
\label{sec:introduction}
\begin{figure*}[ht!]
    \centering
        \includegraphics[width=1\linewidth, trim = 5 5 5 5, clip]
        {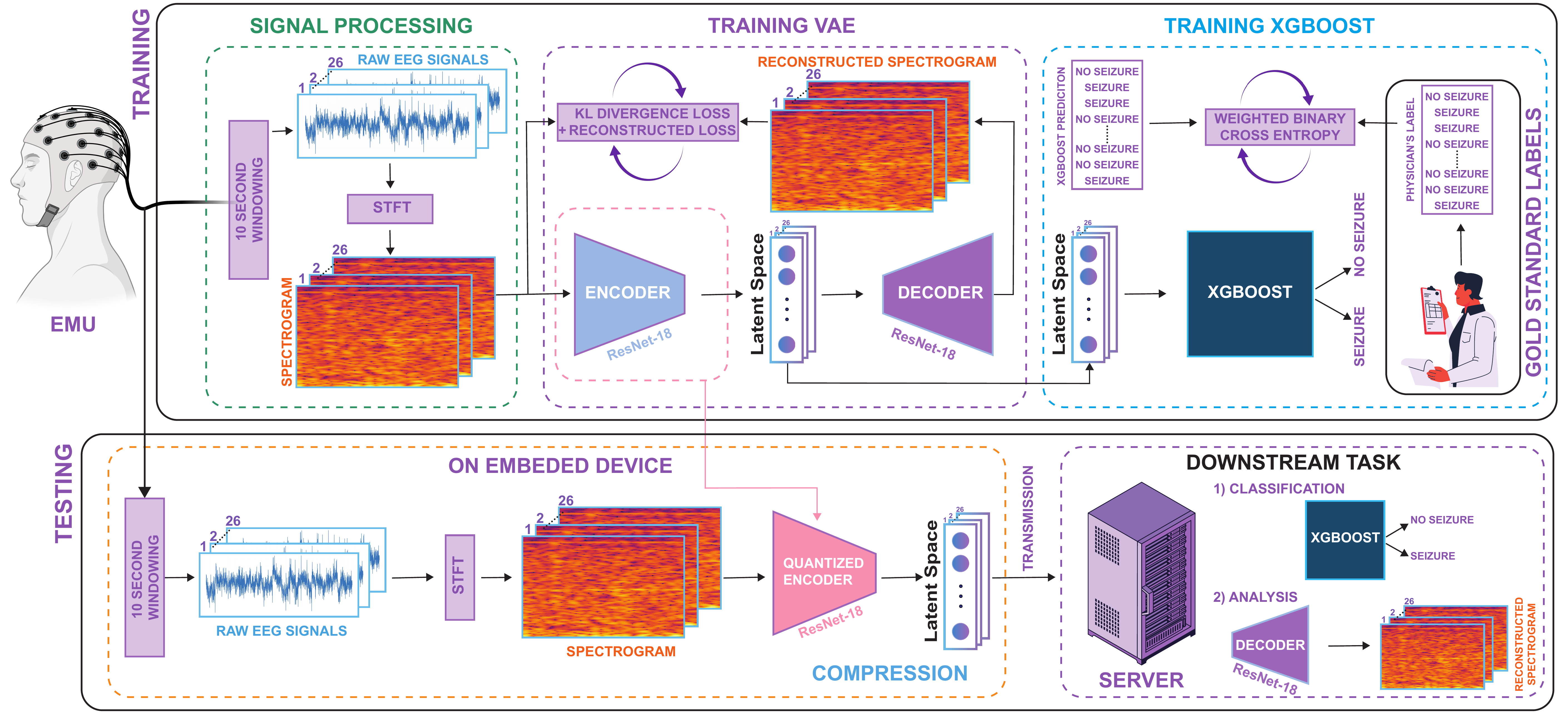}
        \vspace{-20pt}
        \caption{System Overview}
        \vspace{-10pt}
        \label{fig:system overview}
\end{figure*} 

Wearable technologies for healthcare are experiencing significant growth and innovation, with a positive outlook for the market. 
Wearables collect important signals associated with important health events for disease diagnosis and treatment, making the market for wearable medical devices estimated to be worth US 30.1 billion \cite{markte}. 
In addition, the recent development of earables or headband devices makes wearable solutions more attractive and socially acceptable. 
As these devices are worn close to the human head, they can collect important sources of signals (e.g., EEG, EMG, EOG, ECG, HRV, PPG) that are impossible by common form factors such as wrist-worn or arm-worn ~\cite{Wearable1,george2023wearable}. 
For example, EEG signals provide fundamental information about brain states and cognitive function, where analysis of different frequency bands in the EEG can reveal insights into alertness, attention, workload, and emotion.
EMG signals can be used to track electrical activity in muscles, allowing for gesture and expression recognition, control of prosthetics, and understanding of movement intent.
EOG measures eye movements and gaze direction, enabling interfaces that respond to where a user is looking. 
By monitoring these brain and nerve signals non-invasively, wearables may gain insights into cognitive and emotional states, enhancing interfaces and providing biofeedback to improve mindfulness, focus, and well-being. 

While wearable technologies show promise for health monitoring and intuitive interfaces, their widespread adoption faces some challenges. 
Many wearables still have limited battery life, needing daily charging or lacking multi-day use. 
This is partly due to the power required for multiple sensors and data processing \cite{casson2019wearable_battery}. 
Wearables also need strong cross-platform compatibility and integration with phones and apps. 
As wearables incorporate more AI, there are added complexities of algorithm development, clinical validation, and explaining results to users.

To fill this gap, in this paper we propose a new signal compression framework optimized for EEG, EMG, and EOG signals. 
This framework substantially compresses the sensor data while maintaining accuracy for downstream usage. 
Signal compression would reduce power consumption and data transmission needs, enabling longer battery life. 
It also streamlines analysis, requiring less processing power. 
The compression framework is designed to maintain compatibility across platforms. 
It also simplifies AI integration by reducing the data volume fed to algorithms.

However, developing the proposed framework is difficult due to the following challenges:  \textit{\textbf{(1) Compression Algorithm Complexity}}: Balancing compression efficiency with minimal loss of information requires careful algorithm design and optimization from hardware to software optimization. 
\textit{\textbf{(2) Trade-off Between Compression and Accuracy}}: Achieving a balance between compression and maintaining accuracy for downstream tasks can be difficult as aggressive compression might lead to a significant loss of information, negatively impacting the accuracy of tasks relying on the compressed data. 
\textit{\textbf{(3) Model Generalization}}: Ensuring that the compressed signals generalize well to different scenarios, user behaviors, and environments is important for the framework's practical utility, but such investigation has not been done in the past. 
\textit{\textbf{(4) Real-Time Processing}}: If the framework is designed to work in real-time, achieving low-latency signal compression while maintaining accuracy is a challenge, especially when dealing with large datasets. 

To address these challenges, we propose a platform-agnostic variational auto-encoder architecture for reliable and low-complexity compression. In summary, this paper makes the following contributions:

\begin{itemize}[leftmargin=*]
\vspace{-1mm}
    \item We have explored and designed a deep neural network-based compression algorithm to compress the physiological signals such that they retain maximum information.
    \item We have studied the architecture and characteristics of the proposed framework and identified minimal tradeoffs with accuracy.
    \item We have evaluated our system on various edge computing devices like ARM CortexV8, Jetson Nano, etc. to show real-world implementation of our work 
    \item We evaluated the solution through the proposed framework on physiological signals collected on real epileptic seizure patients in the hospital and obtained 91\% accuracy while obtaining a compression ratio of 1:293, and significantly extending the lifetime of the embedded systems. 
\end{itemize}
\section{Background and Motivation}
\label{sec: preliminary}
\begin{figure*}[h!]
    \centering
        \includegraphics[width=1\linewidth]{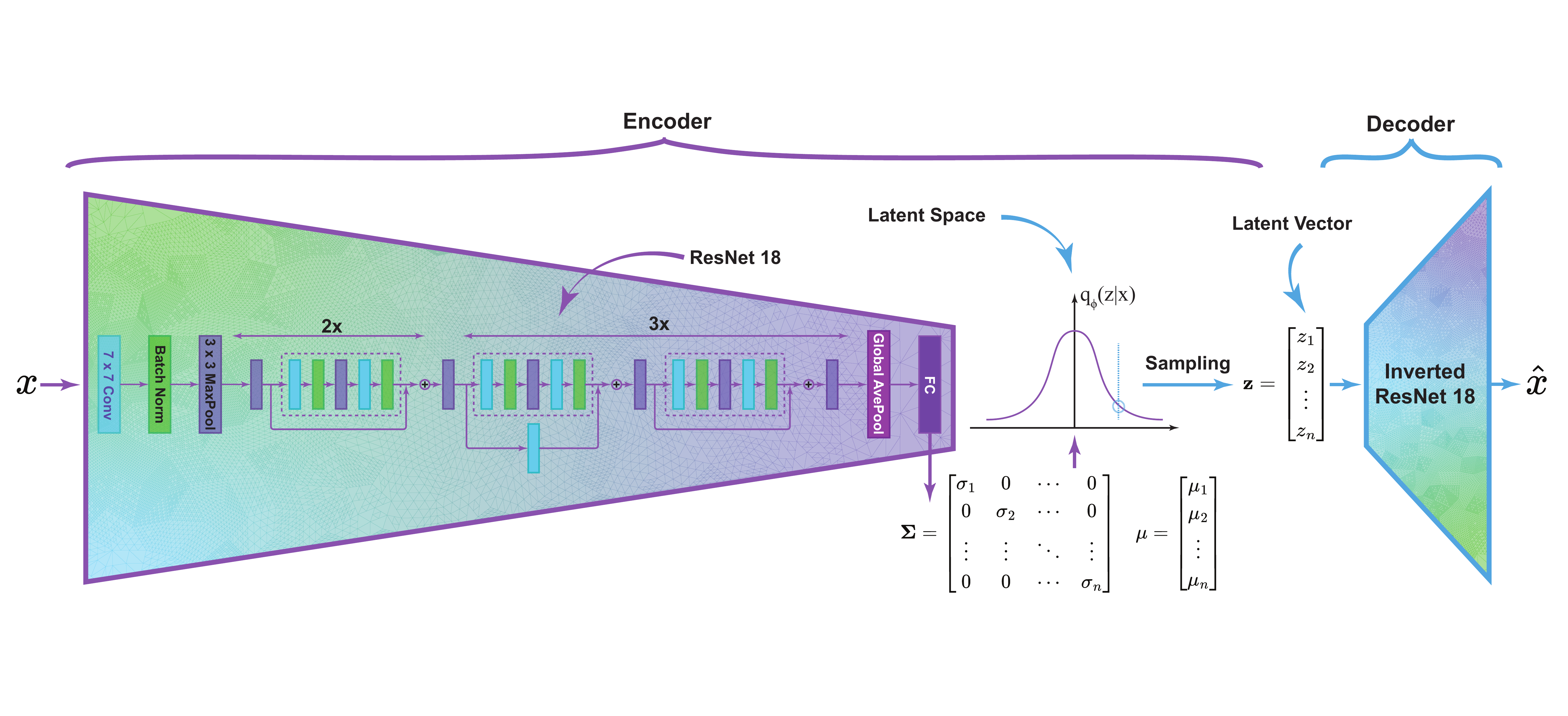}
        \vspace{-55pt}
        \caption{Detailed architecture of the VAE used in this study. Each input is fed through a ResNet 18 network. This network finds the recognition model that has the highest likelihood of generating the input $x$. Then, the decoder (an inverted ResNet18) tries to reconstruct the input based on the $n$ dimensional latent vector learned before.}
        \vspace{-10pt}
        \label{fig:vae}
\end{figure*} 

\textbf{Why Signal Compression?} Compressing signals reduces the size of signal data, making it more efficient to store. 
This is particularly important when dealing with large data, such as sensor data, audio, or video, which can consume significant storage space \cite{arman1993image}. 
 In addition, it helps reduce the amount of data that needs to be transferred among devices and cloud. 
This can lead to faster transmission speeds, lower data transfer costs, and reduced congestion on the network \cite{nicolae2010high}. 
Next, compression can be used as part of data security measures. 
By compressing data before encryption, it may become more resistant to certain types of attack. 
Moreover, for applications where low latency is crucial, compression can reduce the delay introduced during data transmission.

There are existing signal processing-based methods for data/model compression. Some work proposed an efficient preprocessing technique for lossless compression with (2D) matrix arrangement which showed better performance compared to the one-dimensional (1D) compression scheme \cite{srinivasan2010efficient}.
Another lossless compression method is introduced for wireless body sensor networks (WBSNs) to reduce energy consumption based on the discrete wavelet transform (DWT) using the lifting scheme (LS) \cite{azar2018using}. 
Few works present a method to identify P300 event-related potentials (ERPs) in EEG signals using algorithmic clustering based on string compression \cite{sarasa2019algorithmic}. 
Log2 subband compression \cite{rasheed2021lossless} is a technique that compares two bio-medical signals to find the differences between them. 
It achieved low power consumption and fast compression for bio-medical signals. 
Also, the author presented a novel EEG signal compression method based on motion-compensated temporal filtering (MCTF) and discrete wavelet transform (DWT) to exploit spatial redundancy \cite{khalid2020eeg}. Some prior works have explored machine learning for physiological signal compression. For example, one approach used K-means clustering for feature compression and channel ranking to aid signal classification \cite{han2018fast}. Another proposed a semi-supervised learning algorithm for wearable sensor networks, achieving 2.37x compression  \cite{chen2020lossless}. While showing promise, these techniques have not demonstrated edge deployability or reached the high compression ratios we attain. Our approach achieves higher compression with minimal information loss, validated by 91\% seizure detection accuracy nearing original signal performance. Critically, we implement and test our system on embedded devices like Jetson Nano, demonstrating real-world feasibility. By combining exceptional compression with edge-native performance, our machine learning framework significantly advances the state-of-the-art for efficient physiological data analysis.

\vspace{-10pt}
\section{System Overview}
\label{sec:systemOveriew}
In this work, we present a system for compressing and transmitting high-dimensional physiological signals including Electroencephalography (EEG), Electromyography (EMG), Electrooculography (EOG), and electrocardiogram (ECG) 
that retains maximal information content while enabling efficient data transfer with low power consumption. The studies of EGG signals have been investigated across diverse applications, serving both real-time diagnoses~\cite{s21113786,Ekhlasi_2021,Smithii2} and long-term cognitive disease and behavior predictions~\cite{Vicchietti2023,32508975,Bin-1058-9244}. Our approach leverages Variational Auto-Encoder (VAE) \cite{Kingma_2019,kingma2022autoencoding} based architecture implemented on edge devices. VAE are novel in their ability to learn flexible latent representations. Our VAE approach surpasses state-of-the-art performance on compression rate and power efficiency benchmarks. To showcase the practicality of our system we have used in-house data collected under the supervision of medical professionals from real patients suffering from seizures. This validation highlights the system's robustness and real-world utility in the context of healthcare applications.

The research framework, as depicted in Fig \ref{fig:system overview} is structured around two primary phases: the training phase and the testing phase. Within the training phase, a sequence of crucial steps is executed, starting with data processing to ensure the prepared data is conducive to subsequent processing. 
Subsequently, the VAE is trained to perform effective data compression, while the XGBoost algorithm~\cite{Chen_2016_XGboost} is trained for the downstream seizure detection task, thereby validating the reliability of the compressed data. 
Transitioning to the testing phase, we employ quantization techniques \cite{duan2023qarv} to adapt the VAE model for compatibility with edge-device inference. Compressed data is then seamlessly transmitted to server resources.

\subsection{System Training}
Our system training encompasses data preprocessing, VAE-based signal compression, and downstream seizure detection to validate the fidelity of the compressed data.

\textbf{Data Processing.} We gather raw physiological signals from hospital patients, expertly annotated by medical professionals to indicate seizure events. These signals encompass 26 channels, representing data from 26 scalp electrodes at each timestamp. 
To facilitate analysis, we segment the data into 10-second windows for all 26 channels and employ Short-Time Fourier Transform (STFT) \cite{STFT} to convert these windows into spectrograms, capturing vital frequency and temporal information. 
More importantly, each spectrogram is paired with its corresponding seizure label, enabling the training of downstream classification models.

\textbf{Signal Compression.} For signal compression we are using VAE-based (Sec \ref{sec:vae}) architecture.
The input to this VAE model comprises spectrograms derived from temporal segments of EEG signals. 
The VAE framework comprises two fundamental networks: the Encoder and the Decoder. 
The Encoder network generates a compressed representation known as the Latent Space from the input spectrogram. 
In our case, we stack the spectrogram for 26 channels and the encoder will generate latent space for all 26 channels, while the Decoder network is tasked with reconstructing the spectrogram from the latent space created by the Encoder. 
This training process involves comparing the original spectrogram and the spectrogram reconstructed by the VAE for all 26 channels, enabling the model to learn an effective compression mechanism while preserving critical signal information.

\textbf{Demonstration: Head-based Signal Compression in Seizure Detection Application.} We selected seizure detection as our demonstration application because it is a widely recognized neurological disorder where collecting multiple physiological signals from the head region can provide crucial insights \cite{compression_classification}. 
The ability to effectively monitor and analyze physiological signals is extremely useful for identifying the onset of seizures and guiding treatment. 
We therefore employ seizure detection as a downstream task to assess the fidelity of the compressed data using the latent space (Sec \ref{sec:xgboost}). 
To facilitate this, we transform our problem into a supervised two-class binary classification, distinguishing between seizure and non-seizure events, utilizing the XGBoost algorithm \cite{Chen_2016_XGboost}. The latent space, generated by the encoder network of the trained VAE, is provided as input to XGBoost for the classification task. Labels provided by medical professionals serve as the training data for the XGBoost model, enabling it to effectively discern and classify these critical events, thus validating the reliability of our compressed data.

\vspace{-10pt}
\subsection{On Chip Testing}
To demonstrate the scalability and practicality of our system, we deploy and evaluate it on common resource-constrained edge devices including ARM CortexV8 and Jetson Nano boards. 
Deploying ML capabilities directly on edge devices enables real-time intelligent analysis with low latency, efficiently processing data at the source before transmission, and preserving privacy by limiting external connectivity. 

\textbf{On-Embedded Chip Compression.} Similar to the training phase, raw EEG signals are converted into 10-second segments of a spectrogram for all 26 channels. 
These raw spectrograms are sent to the VAE for further processing. 
To make the VAE encoder compatible with edge devices we perform quantization of VAE (Sec: \ref{subsec:quantization}), specifically we quantized the encoder network of the VAE and infer it on the chip. 
The stacked spectrograms are sent to the quantized encoder of the VAE which gives stacked latent space for 26 channels. 
This latent space is transmitted from the chip to the server for downstream tasks.

\textbf{Downstream on Server.} The Transferred latent space is then utilized on the server for downstream tasks, which include \textbf{classification} and \textbf{analysis}. For classification, we used our trained XGBoost model to detect seizures, while for analysis we used the trained decoder network of the VAE to reconstruct the raw spectrograms which can be further used for expert analysis, thus showing the practical implementation of the entire system.

\section{Variational Auto-Encoder}
\label{sec:vae}

\begin{figure}[t!]
    \includegraphics[width=1\linewidth]{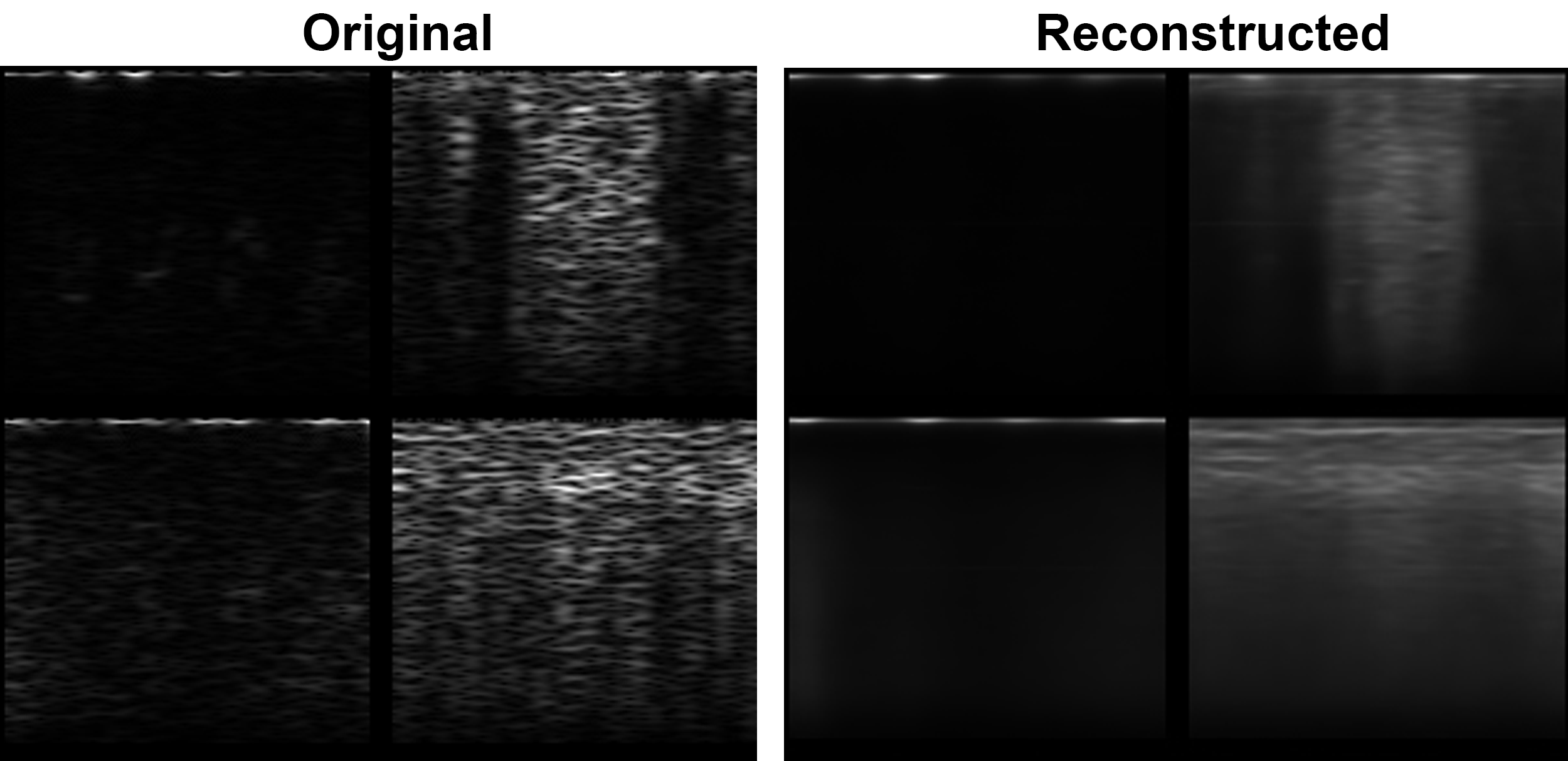}
    \vspace{-20pt}
    \caption{Four randomly selected samples from the test set and their reconstructed versions when latent size is 64. Note that due to normalization and gray-scale transformation, the samples are not in color.}
    \label{fig:reconstructed}
    \vspace{-10pt}
\end{figure}

\begin{figure}[t!]
    \includegraphics[width=1\linewidth,trim = 20 15 20 20, clip ]{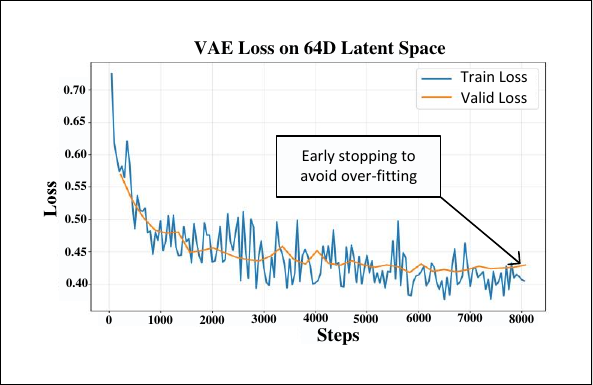}
    \vspace{-20pt}
    \caption{Training and validation loss (KL divergence loss plus reconstruction loss) graph for VAE for 64-dimension latent space over steps (i.e. iteration on different batchs)}
    \label{fig:vaeloss}
      \vspace{-10pt}
\end{figure}

Autoencoders (AEs) are artificial neural networks designed to efficiently represent data by mapping inputs to a fixed-size, usually low-dimensional latent space, and then reconstructing the original data from this representation. Comprising two main components, the encoder and the decoder (Fig. \ref{fig:vae}), the encoder compresses the input, and the decoder reconstructs it from the latent space \cite{bank_autoencoders_2020}. 
 The network is trained to minimize the difference between the input and the reconstructed output, enabling applications in dimensionality reduction \cite{wang_vasc_2018}, anomaly detection \cite{sun_learning_2018}, medical image processing \cite{nasr_clinically_2023, hajighasemi_multimodal_2023} and more.

Variational Autoencoders (VAEs) are a special case of AEs where the network tries to uncover the underlying distribution associated with these latent variables instead of only finding the latent representation of individual data points. Once the latent distributions are learned, latent representations can be simply sampled from each distribution.
In fact, VAE is a type of generative model that can effectively compress signals by mapping them to a lower-dimensional latent space while preserving their essential characteristics. Our VAE encoder maps multi-channel physiological signals into compact latent representations. The decoder reconstructs the original signals from these compressed latent vectors. 
By training the encoder and decoder together, the VAE learns to compress the signals substantially while retaining critical information. This allows efficient analysis of physiological data even after high compression.

To achieve this, the model assumes the observed (training) dataset $ \boldsymbol{X} = \left\{x^{\left(i\right)}\right\}_{i=1}^{N} $ consists of $N$ independent and identically distributed derived from some random variable $\boldsymbol{x}$. More importantly, $\boldsymbol{x}$ itself is also the result of a random process involving an unseen random variable $\boldsymbol{z}$ (latent variable). In mathematical language, $x^{\left( i \right)} \sim p_{\theta^{*}}\left( \boldsymbol{x} | \boldsymbol{z} = z^{\left( i \right)} \right)$ where $z^{\left( i \right)} \sim p_{\theta^{*}}\left( \boldsymbol{z} \right)$. The other underlying assumption here is that both $\boldsymbol{z}$ and $\boldsymbol{x}$ come from the same family of parametric distributions $p_\theta$ with the true parameters $\boldsymbol{\theta^*}$ \cite{kingma_auto-encoding_2022}. The goal is to find $\boldsymbol{\theta^*}$ based on the observed dataset.

If $\boldsymbol{\theta}$ is a candidate solution, according to the Bayes rule (Eq. \ref{eq:bayes}), we can summarize all of our prior assumptions and prior knowledge about the latent distribution given the choice of candidate parameters in the prior (i.e. $p_{\boldsymbol{\theta}}\left( \boldsymbol{z} \right)$). The posterior (i.e. $p_{\boldsymbol{\theta}}\left( \boldsymbol{z} | \boldsymbol{x} \right)$), actually tells us how to update our belief about the latent distribution once we take into account the reported observations $\boldsymbol{X}$. Once our beliefs are updated, we have no use for our prior knowledge, and thus, we can consider the posterior as the new prior ($p_{\boldsymbol{\theta}}\left( \boldsymbol{z} | \boldsymbol{x} \right) \rightarrow p_{\boldsymbol{\theta}}\left( \boldsymbol{z} \right)$). This is the essence of Bayesian inference.

Hence, by finding the posterior and subsequently the new prior, for a sample point $x^{\left( i \right)}$ we can find the latent representation of it by sampling from the posterior $ z^{(i)} \sim p_{\boldsymbol{\theta}}\left( \boldsymbol{z} | \boldsymbol{x} = x^{\left( i \right)} \right)$. Moreover, since calculating the posterior entails finding the likelihood (i.e. $p_{\boldsymbol{\theta}}\left( \boldsymbol{x} | \boldsymbol{z} \right)$), as a bonus, we can sample from the likelihood $ \hat{x}^{(i)} \sim p_{\boldsymbol{\theta}}\left( \boldsymbol{x} | \boldsymbol{z} = z^{\left( i \right)} \right)$ to generate fake samples that resemble $x^{\left( i \right)}$.

\begin{equation}
    p_{\boldsymbol{\theta}}\left( \boldsymbol{z} | \boldsymbol{x} \right) = {p_{\boldsymbol{\theta}}\left( \boldsymbol{x} | \boldsymbol{z} \right) p_{\boldsymbol{\theta}}\left( \boldsymbol{z} \right)} / {p_{\boldsymbol{\theta}}\left( \boldsymbol{x} \right)}
\label{eq:bayes}
\end{equation}

However, the marginal likelihood (i.e. $p_{\boldsymbol{\theta}}\left( \boldsymbol{x} \right)$), and therefore the posterior, are not tractable in most cases since it requires integration over the set of all possible choices for latent variables $\boldsymbol{\theta}$ ($p_{\boldsymbol{\theta}}\left( \boldsymbol{x} \right) = \int_{z} {p_{\boldsymbol{\theta}}\left( \boldsymbol{x} | z \right) p_{\boldsymbol{\theta}}\left( z \right)}$). To overcome this challenge, a recognition model $q_{\boldsymbol{\phi}} \left( \boldsymbol{z} | \boldsymbol{x} \right)$ is introduced as an approximation of the true posterior $p_{\boldsymbol{\theta}}\left( \boldsymbol{z} | \boldsymbol{x} \right)$. The power of VAEs comes from being able to solve for both generative parameters $\boldsymbol{\theta}$ and variational parameters $\boldsymbol{\phi}$ simultaneously using deep neural networks.

In VAEs, the encoder is a deep network that tries to approximate the recognition model $q_{\boldsymbol{\phi}} \left( \boldsymbol{z} | \boldsymbol{x} \right)$. It takes in a data point $x^{\left( i \right)}$ and finds the parameters of a distribution over $\boldsymbol{z}$ from which $x^{\left( i \right)}$ could have been created. The decoder is also a deep network that approximates the maximum likelihood $\max\{p_{\boldsymbol{\theta}}\left( \boldsymbol{x} | \boldsymbol{z} \right)\}$. It takes in a random sample from the distribution learned by the encoder and finds $\hat{x}^{\left( i \right)}$ that has the highest likelihood to be have been generated from the sampled latent vector (Fig. \ref{fig:vae}). 

It can be shown that if the sum of negative variational lower bounds ($\mathcal{L}$ in Eq. \ref{eq:loss}) over the data point is minimized, the encoder and decoder networks are trained by the Bayesian inference paradigm \cite{kingma_auto-encoding_2022}. This would be a loss to train the network with. The first term in the loss is the KL divergence between the recognition model and the prior. KL divergence is a measure that quantifies the difference between two probability distributions \cite{kullback_information_1951}. The second term is however the negative expected reconstruction error. In the loss function, the second term tries to make sure that, on average, the reconstructed data has the highest likelihood of being similar to the input data, whilst the second term regularizes the network and does not allow it to learn overly complicated distributions by making sure the approximate posterior remains as similar as possible to the prior.
\vspace{-5pt}
\begin{align}
    - \mathcal{L} \left(\boldsymbol\theta, \boldsymbol\phi; \boldsymbol{x^{(i)}} \right) = & D_{KL} \left( q_{\boldsymbol\phi} \left( \boldsymbol{z}|\boldsymbol{x^{(i)}} || p_{\boldsymbol\theta} \left( \boldsymbol{z} \right)\right) \right) \nonumber \\
    & - \mathbb{E}_{q_{\boldsymbol\phi} \left( \boldsymbol{z}|\boldsymbol{x^{(i)}} \right)} \left[ \log{p_\theta \left( \boldsymbol{x^{(i)}} | \boldsymbol{z} \right)} \right] \nonumber \\
    & = \mathcal{L}_{{KL}}^{(i)} + \mathcal{L}_{{Recon}}^{(i)}
\label{eq:loss}
\end{align}

For this study, we chose the before be the centered isotropic multivariate Gaussian $p_{\boldsymbol{\theta}}\left( \boldsymbol{z} \right) \sim \mathcal{N}\left(\boldsymbol{0}, \boldsymbol{I} \right)$, and that $p_{\boldsymbol{\theta}}\left( \boldsymbol{x} | \boldsymbol{z} \right) \sim \mathcal{N}\left(\boldsymbol{\mu}, \boldsymbol{\Sigma} \right)$ is also a multivariate Gaussian with parameters $\boldsymbol{\mu}$ and $\boldsymbol{\Sigma}$ that are learned by the encoder. Moreover, the networks used inside the encoder and decoder are ResNet 18 and inverted ResNet 18, respectively (Fig. \ref{fig:vae}). ResNets, or Residual Networks, are convolutional neural network architectures that introduce skip connections, or shortcuts, to jump over some layers, enabling the training of much deeper networks by mitigating the vanishing gradient problem \cite{he_deep_2015}.

\section{VAE on the Chip}
\label{sec:vaeOnChip}

In the context of modern computational research, the efficiency and adaptability of systems across varying hardware architectures are of extreme importance. This research aims to evaluate the performance and practicality of our system on edge devices. 
Central to this is the application of quantization on edge devices, a technique critical to the optimization of our model performance. 
This section characterizes the system's hardware and architecture that we will use for our purpose, and then we will talk about the quantization of the compression model to implement the system on edge device.

\vspace{-10pt}
\subsection{Architecture}

In this discussion, we use Raspherry Pi and Jetson Nano as examples to discuss the common structure of two famous commercial edge products. The popularity of these two devices is due to their mobility, cost, and customizability since they are both development boards.

\textbf{Raspberry Pi} has introduced many interesting capabilities into the world of computing. 
There are two specific ways it has evolved the field of computing, those being mobility and size. It can act as a normal desktop computer that is about the size of a phone. Due to its small size, this makes it an ideal platform for mobile applications. Because of this, it has become a very popular mobile computing platform. The Raspberry Pi 4 Model B hosts a Broadcom ARMV8 Cortex-A72 64-bit Quad Core processor with 4GB SDRAM with a maximum power draw of 15W \cite{raspberrypiRaspberryModel}. The Cortex-A72 boasts a 3.5x performance increase and 75\% energy efficiency on target processes over the previous Cortex-A15 \cite{7482541}. This is a massive leap above previous ARM chips and works greatly to our benefit in concern of our compressed model.

\textbf{Jetson Nano} took the things that made the Raspberry Pi great and improved upon it. The Nano is only slightly larger than a Raspberry Pi, but its slightly larger size is due to 
having a designated graphics card built into it. This has made it an ideal mobile computing platform for Artificial Intelligence and Machine Learning applications \cite{9126102,9152915}. Because of its dedicated GPU (graphics processing unit), it is able to handle applications involving AI and Machine Learning with much more ease than the Raspberry Pi can.

For this project, the compression model is able to run on the Raspberry Pi, Jetson Nano and were able to make it utilize edge computing by being able to send the compressed output back to the Nvidia DGX A-100 Server. Due to its extreme mobility and ease of ability to perform the calculations necessary for this project, this makes the Jetson Nano and Raspberry Pi the ideal choice of implementation for the end-user edge computing hardware \cite{9610432}.

\subsection{Quantization of VAE}
\label{subsec:quantization}

In the past few years, there has been a shift towards edge computing. A whopping 45\% of data will be processed on an edge network \cite{9083958}. 
This is where quantization comes into play, allowing us the opportunity to run complex models on large data efficiently on edge devices. 

Edge-computing devices, like the Raspberry Pi 4 and Jetson Nano (Sec \ref{sec:vaeOnChip}) and other ARM Cortex chips(i.e., common platforms for edge-computing research), are revolutionizing the way we process data closer to its source. 
These devices, while compact and cost-effective bring forth a problem of optimizing computational tasks, especially when dealing with complex neural networks such as our VAE model. 
In essence, quantization takes continuous infinite values and maps them onto a smaller set of discrete finite values. 
This in terms of edge devices can allow for the approximation of weights and place certain limits, such as bit-precision and range of the values. 
When used alongside the Raspberry Pi and Jetson Nano, quantization gives us the ability to represent data with a reduced number of bits within our VAE model with minimal compromise on model performance.

The quantization approach we adopted for our needs involved post-training quantization. Our research has led us to modify the weights of our on-chip VAE model to a lower bit precision. The size of our VAE model before undergoing quantization was 400MB and the size of the encoder was 200MB. Post-training quantization led us to modify the bit-precision of our model from float64 down to float16. This greatly reduced the overall size of the model and ensured it could fit on the Raspberry Pi or Jetson Nano without any problems. After post-training quantization, the overall size of the VAE model was 150MB with the encoder being 75MB.

In summary, quantization enables complex neural network models like our VAE to be efficiently deployed on resource-constrained edge devices. Through post-training quantization, we reduced the bit precision of our model, maintaining performance, decreasing the model size to fit onboard edge hardware like Raspberry Pi and Jetson Nano.



\section{XGBoost}
\label{sec:xgboost}

In this study, we embrace a two-fold approach to EEG signal classification, driven by the utilization and compressed latent space vectors. The core motivation behind this approach is to underscore the value of data compression through the application of Variational Autoencoders (VAEs), while showcasing the adaptability and robustness XGBoost classifiers in different classification scenarios.
In fact, XGBoost (Extreme Gradient Boosting) is a popular machine learning algorithm that belongs to the family of ensemble learning methods. It's designed for both classification and regression tasks and has gained significant attention and success in various data science competitions and real-world applications due to its high predictive accuracy and efficiency. It builds an ensemble of decision trees, where each tree is trained to correct the errors of the previous tree in the sequence and combines the predictions of multiple weak learners (individual decision trees) to create a stronger final prediction.

To explore the effectiveness of data compression and its implications for classification, we used the concept of latent space representation using VAEs. Variational Autoencoders allow us to distill the most salient features of the EEG data into a reduced-dimensional latent space, retaining essential information while discarding noise and redundancies. We opt for the XGBoost classifier due to its versatility and interpretability. XGBoost, a gradient-boosting algorithm, excels in capturing complex relationships in structured data with regularized latent space, making it an ideal choice for compressed latent vectors and enabling effective classification.

The classifier was tuned with parameters that optimize its performance. It was configured with 500 estimators, fostering robustness through a more extensive ensemble. The maximum tree depth of 6 facilitated capturing intricate relationships within the data, while the lower learning rate of 0.01 allowed for precise adjustments during training. To mitigate class imbalance, we strategically adjusted the weight parameter to reflect the varying class frequencies, thus enhancing classification accuracy. Additionally, regularization was applied through the reg\_lambda and reg\_alpha parameters, aiding in preventing overfitting. Employing early stopping with 50 rounds allowed us to avoid overfitting while monitoring validation accuracy during training. To assess the model's performance, we employed a 10-fold cross-validation strategy, ensuring its generalization capabilities across diverse subsets of the data. Through meticulous parameter tuning and a rigorous validation approach, we demonstrate the XGBoost classifier's capability to accurately classify signals within the compressed latent space.

\section{A Real-world Study}
\label{sec:realworld}

\subsection{Data Acquisition}
\label{subsec: Datacolection}
In medical practice, the primary way in which doctors diagnose neurological disorders is through Electroencephalogram (EEG) tests. 
However, this diagnosis method requires trained technicians to set up and operate the test device which means patients must be hospitalized making this an expensive process. 
Patients are generally admitted into the neurology department of hospitals for week-long studies and made to wear EEG headcaps which usually have more than 20 wired electrodes. 
Neurological experts then analyze the resulting signals to determine characteristic patterns of illnesses such as epileptic seizures, Alzheimer's, and brain tumors to prescribe the next course of treatment. 
The EEG test is considered the gold-standard test for such diagnoses \cite{shellhaas_continuous_2015} thus, improving the testing methodology is an important research focus. 

For our work on compressing and transmitting the high-dimensional physiological signals collected from the EEG device, we conducted a study on real patients admitted to the Epileptic Monitoring Unit (EMU) of a hospital for their seizure diagnosis test. 
Through our collaboration with the hospital, we gained access to patients admitted to the EMU for long-term EEG monitoring. 
In order to be eligible for our study, participants were required to be at least 18 years old at the time of enrollment. 
Similarly, patients who were either unable or unwilling to provide informed consent were also excluded from our analysis. Following these criteria, we successfully recruited a total of 33 patients, between ages 19 to 74. The gender distribution within the sample included 53.1\% biological males and 46.9\% biological females. It should be mentioned that before beginning the patient study, we obtained approval from the Institutional Review Board (IRB). During our study, the patients were continuously monitored in their rooms following standard clinical practices. At the end, all data that was recorded were deidentified, except for the patient's study ID. The EEG data was subsequently stored in an encrypted and password-protected laptop for processing before being securely stored on the REDCap (Research Electronic Data Capture) Database \cite{noauthor_about_nodate}. All local data on the laptop was destroyed following its transfer to REDCap. The data collected from the participants was completely anonymized and was exclusively shared among the researchers of the study.

\noindent \textbf{The Dataset.} The EMU doctors then examined the recordings and provided us with the onset and offset times of each seizure. A portion of our dataset is shown in table \ref{table:dataset}. 
\begin{table}[!t]
    \centering
    \begin{tabular}{|c|l|l|l|}
    \hline
        \textbf{Patient ID} & \textbf{Start Time} & \textbf{End Time} \\ 
        \hline
        P1 & 21:42:56 & 21:43:21 \\ 
        \hline
        P2 & 01:01:01 & 01:01:58 \\
        \hline
        P3 & 15:01:49 & 15:02:55 \\ 
        \hline
        P4 & 22:55:36 & 22:56:53 \\
        \hline
        P5 & 14:37:08 & 14:38:04 \\
        \hline
    \end{tabular}
    \caption{Seizure log labeled by doctors}
    \vspace{-30pt}
    \label{table:dataset}
\end{table}
There are some inherent complexities when labeling the onset and offset times. The onset often occurs very early on in a seizure when the symptoms are faint, so it is not easily identified. Conversely, at the later stages, a lot of external noise is seen in the EEG signals as medical personnel enter the room at this point to attend to the patient, performing various activities which distort the signal making it challenging to find the precise offset point. Thus we excluded the events where doctors could not confidently label the timestamps to ensure the reliability of our dataset. 

From our data collection step, we were able to collect a total of 14 events from the 7 patients who experienced seizures during our study; the rest did not experience an attack while in the hospital. 
The American Clinical Neurophysiology Society categorizes an abnormal event as a seizure if it lasts for at least 10 seconds \cite{hirsch_american_2021}. Following this, we split our signals into 10-second segments. If a 10-second segment's start and end time fell entirely within the timestamp of a seizure event, we considered that sample as a seizure. If a segment's start or end time partially overlapped the seizure timestamp or did not overlap at all, we labeled it as a non-seizure sample. A seizure scenario follows a pattern of non-seizure times followed by the seizure period and then the seizure ends to produce more non-seizure times. To cover this scenario, we included some samples marked as non-seizure from the pre-seizure stage, followed by the samples marked as seizures, and then some more non-seizure samples from the post-seizure stage. Together, these segments constituted one single event in our resultant dataset. Since we had only 14 events in our dataset, we augmented the dataset using strides and sliding windows. We used a 1-second stride where we slid the window 1 second forward meaning there was a 9-second overlap between the successive windows which provided us with the maximum number of data points. For each window, we apply a Short-time Fourier Transform (STFT) to convert the raw signals from all 26 electrodes into corresponding spectrogram representations. Spectrograms effectively capture the temporal dynamics and spatial correlations across channels in a compact time-frequency representation. This makes them well-suited for training machine learning models /
We labeled each set of 26 spectrograms per window based on timestamps provided by medical experts, thus providing us with an extensive labeled dataset for our machine-learning algorithm.
\begin{figure}[t]
    \centering
    \includegraphics[width=0.5\textwidth,trim=20 10 10 5,clip]{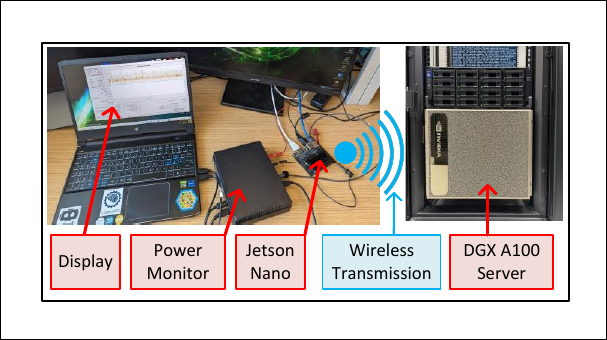}
    \vspace{-30pt}
    \caption{Hardware setup for power measurement and data transfer. Quantized VAE used on Jetson Nano. Power measure for both compression and transfer of compressed data to the server.}
    \label{fig:Hardware_setup}
    \vspace{-15pt}
\end{figure}

\begin{figure*}[t!]
    \centering
        \includegraphics[width=1\linewidth, trim = 10 5 5 5, clip]{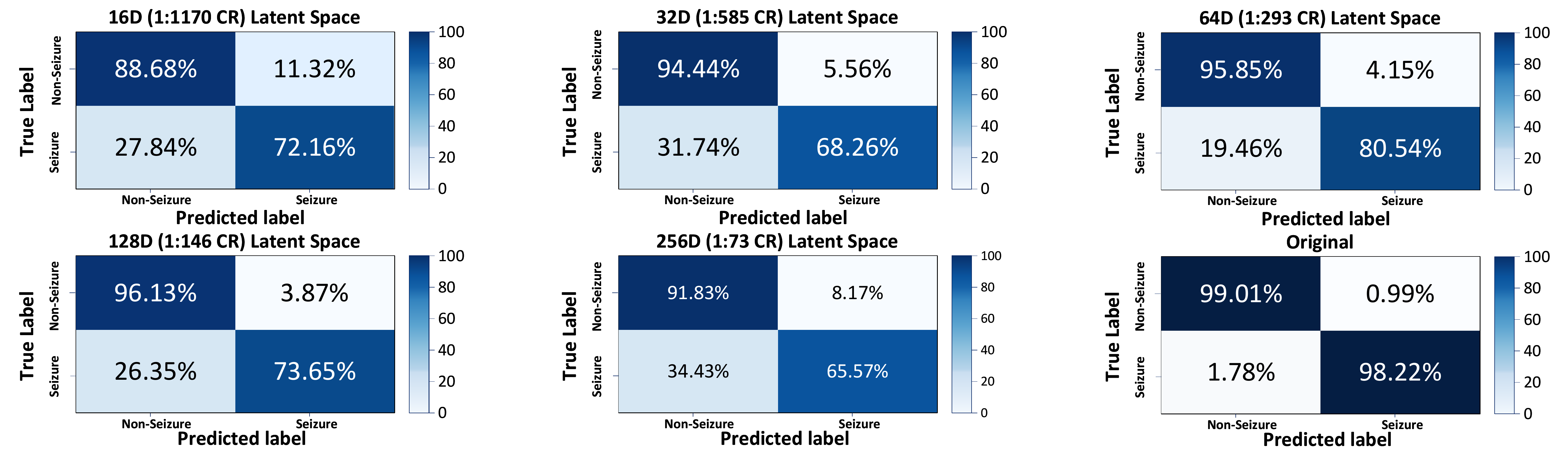}
        \vspace{-10pt}
        \caption{Confusion matrix for XGBoost's performance for detecting seizure from different dimension's latent space of the compressed signals }
        \label{fig: confusionmatrix}
        \vspace{-15pt}
\end{figure*}

\subsection{Hardware/Software Implementation}
\textbf{On the chip} To evaluate the performance of the proposed framework, we conduct the following experiments. 
These devices receive the multi-channel raw EEG signals and transform them into spectrogram representations capturing temporal and frequency information \cite{Aslan_Akin_2020}. The on-chip encoder network compresses each input spectrogram into a compact latent space vector. Performing this high-dimensional data encoding locally reduces transmission bandwidth and latency compared to sending raw EEG readings. The encoder leverages efficient neural network operators and quantization to enable real-time data compression under the constrained memory and compute of edge devices

\textbf{Server Implementation}
Once they have compressed the data, it can be transmitted to the Nvidia DGX A-100 server via Secure Copy Protocol (SCP), and more computationally difficult downstream tasks are completed on the server. With the reduced data size, the transfer should take less time and resources than it would otherwise. Once received by the server, compressed signal/latent space is sent to the classification model for seizure detection or to the decoder network to reconstruct the spectrogram for expert analysis

\section{Performance Evaluation}
\label{sec:result}

In this section, we evaluate the performance of our proposed framework through comprehensive real-world experiments. We assess the VAE compression capabilities, classification accuracy, power consumption, and medical significance based on metrics including reconstruction loss, precision-recall curves, confusion matrix, energy expenditure, and potential clinical benefits.

\subsection{VAE Performance}
As was discussed in Sec \ref{sec:vae}, the loss function of the VAE is comprised of two parts: KL loss and reconstruction loss. KL loss plays a regularizer role to avoid overfitting, while the reconstruction loss makes sure that the reconstructed output has a high likelihood of resembling the input data. Since we control the overfitting by employing early stopping, and we are mainly focused on compressing the input as opposed to generating fake outputs, we have decided to train the models using a weighted version of the loss as presented in Eq. \ref{eq:weighted_loss}. Here, $\lambda$ gauges the trade-off between compression and overfitting. For all of our experiments, we used $\lambda=0.1$.
\begin{equation}
    \boldsymbol{L}_{total} = - \sum_{i = 1}^N \left( \lambda \mathcal{L}_{{KL}}^{(i)} + \mathcal{L}_{{Recon}}^{(i)}\right)
\label{eq:weighted_loss}
\end{equation}

In Fig \ref{fig:vaeloss}, you can see the change of $\boldsymbol{L}_{total}$ for both the training set and the validation set as the model is being trained for a latent size of 64. We used an early stopping with a patience of 10 monitoring the validation total loss. The result is that we make sure we stop the training before the model overfits the data. We used two Nvidia A100 80 GB GPUs on a DGX workstation to train all models using a batch size of 128. We also used an ADAM optimizer with a learning of 0.0001 to train all models.

Although the reconstructed spectrograms are of no use for our purpose of siezure detection, just to showcase that even at high compression ratios the latent space still preserves enough information about the input, we have illustrated four random test images and their corresponding reconstructed version when the latent size is 64 in Fig. \ref{fig:reconstructed}.

\begin{figure*}[ht!]
    \centering
        \includegraphics[width=1\linewidth, trim = 5 5 5 5, clip]{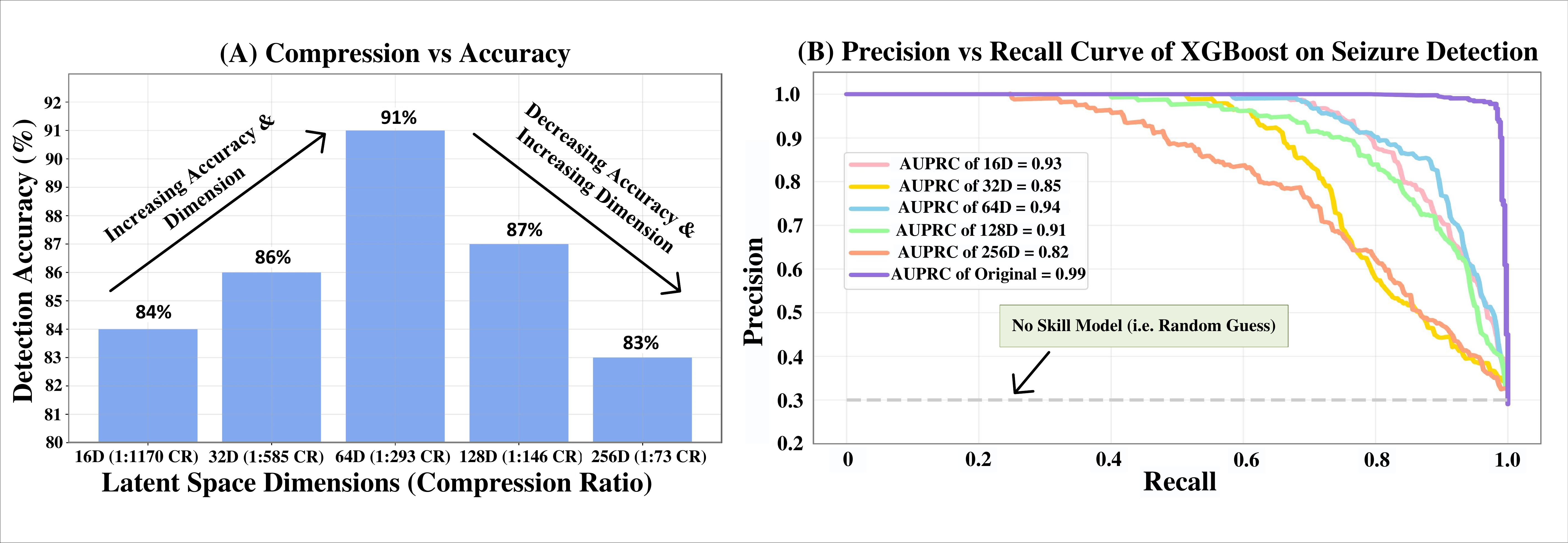}
        \vspace{-30pt}
        \caption{(A) shows the accuracy of different compression ratios and (B) shows Precision-Recall curve for different compression ratios}
        \label{fig: Classification 1}
         \vspace{-10pt}
\end{figure*}

\subsection{Classification Performance}

We evaluate our model from three perspectives: overall accuracy, confusion matrix analysis, and precision-recall curves. This provides comprehensive insights into performance.

Our multi-faceted analysis consistently shows the 64-sized latent space performs excellently. It achieves 91\% accuracy, balanced true positives/negatives in the confusion matrix, and the best precision-recall AUC despite data imbalance. By evaluating across diverse metrics, we gain a holistic view of the model's effectiveness. The 64-sized latent space demonstrates consistently strong capabilities.

\textbf{Accuracy}
As shown in fig \ref{fig: Classification 1}A the accuracy plot provides a comprehensive overview of how varying latent vector dimensions impact the model's ability to distinguish between seizure and non-seizure instances. 
 For extensive validation, we evaluated our system using the K-fold cross-validation technique (K=10) among the considered latent vector sizes (16, 32, 64, 128, 256), the 64-sized latent vectors stand out as particularly effective, achieving a substantial accuracy rate of 91 percent. This suggests that this specific latent space configuration captures essential features that significantly contribute to accurate seizure classification.

\textbf{Confusion Matrix} 
In this sub-subsection, we provide the confusion matrix that illustrates a comprehensive evaluation of model's performance in classifying seizure and not-seizure cases. In general, confusion matrix presents a breakdown of predicted classification against actual class labels and provide us with four values including True Positive, False Positive, True Negative, and False Negative. With these values, we can calculate Recall, Precision, Specificity, and Accuracy which show a concise and insightful assessment of the models' efficiency in categorizing data. 

Fig \ref{fig: confusionmatrix} shows the confusion matrix related to our model for five different latent space sizes (16, 32, 64, 128, 256) and the original one. In order to compare the results we need to find the latent size which has better accuracy for both True Positive and True Negative. Among them 128D and 64D have better accuracy for classifying non-seizure cases in comparison with others with the accuracy of 96.13 and 95.85 percent respectively. However 64D latent size is more accurate in seizure instances compare with 128D latent size with the accuracy of 80.54 vs 73.65 percent. Thus, among these latent spaces the 64-sized one outperforms others in classifying the instances. 

\textbf{PR-Curve}
In this sub-subsection, we delve into the Precision-Recall (PR) curves that assess the classification performance for seizure vs. not-seizure instances across multiple latent spaces (16, 32, 64, 128, 256) as well as the original data. PR curves provide a valuable perspective on the model's precision and recall trade-offs, especially crucial when dealing with imbalanced datasets like ours.

Precision and recall have distinct interpretations in the context of PR curves. The horizontal axis of the PR curve represents recall, illustrating the model's ability to capture positive instances from the actual positives. Higher recall means more true positives are identified. The vertical axis, on the other hand, signifies precision, showcasing the model's precision in correctly classifying predicted positives. Higher precision means fewer false positives.
precision measures the accuracy of positive predictions, while recall quantifies the model's capacity to find all actual positives.

\begin{equation}
\text{Precision} = \frac{\text{True Positives}}{\text{True Positives} + \text{False Positives}}
\end{equation}

\begin{equation}
\text{Recall} = \frac{\text{True Positives}}{\text{True Positives} + \text{False Negatives}}
\end{equation}

\begin{figure*}[ht!]
    \centering
        \includegraphics[width=1\linewidth, trim = 5 5 5 5, clip]{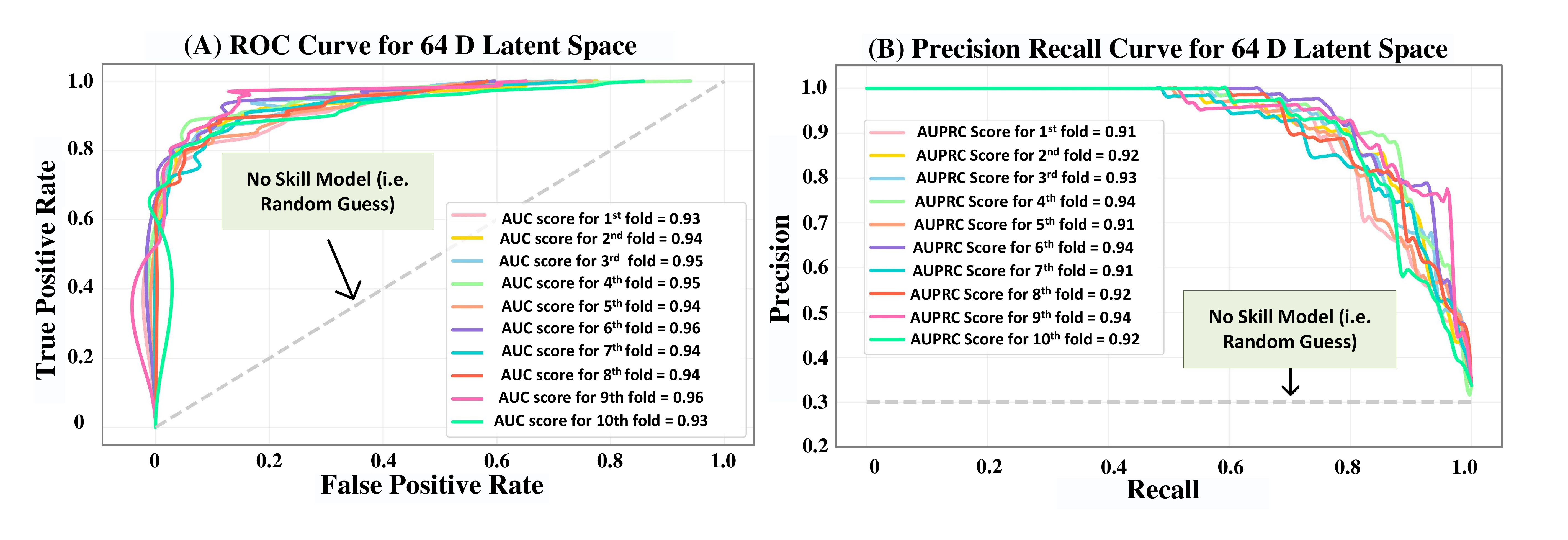}
        \vspace{-30pt}
        \caption{(A) ROC curve for XGBoost performance on 64D latent space for seizure detection. (B) Precision-Recall curve for XGBoost performance on 64D latent space for seizure detection. }
        \label{fig: classification_kfold}
        \vspace{-10pt}
\end{figure*}

In our study, fig \ref{fig: Classification 1}B shows the PR-AUC values for the original data and latent spaces were as follows: 0.99 (original), 0.85 (16), 0.85 (32), 0.94 (64), 0.91 (128), and 0.82 (256). Notably, the 64-sized latent space exhibited the highest PR-AUC, indicating a commendable balance between precision and recall. This suggests that the 64-sized latent space encapsulates pertinent information necessary for accurate seizure classification. Also, we implemented the K-fold cross-validation technique (K=10) for 64-sized latent and created the Precision-Recall curve for that which fig \ref{fig: classification_kfold}B shows.

\textbf{ROC-Curve} (Receiver Operating Characteristic) curve is a graphical representation commonly used in binary classification tasks to illustrate the performance of a classification model. It displays the relationship between the true positive rate (sensitivity) and the false positive rate (1-specificity) across different thresholds used to classify instances. A good classifier's ROC curve should be positioned as close as possible to the top-left corner of the plot, which represents high sensitivity (true positive rate) and low false positive rate. The area under the ROC curve (AUC-ROC) is often used as a single metric to summarize the overall performance of a classification model. AUC-ROC ranges from 0 to 1, where a value closer to 1 indicates better performance.
we used our K-fold cross-validation technique for 64-sized latent to create ROC curve which fig \ref{fig: classification_kfold}A shows.

\begin{figure}[t]
\centering
\includegraphics[width=0.5\textwidth,trim = 40 30 150 30, clip]{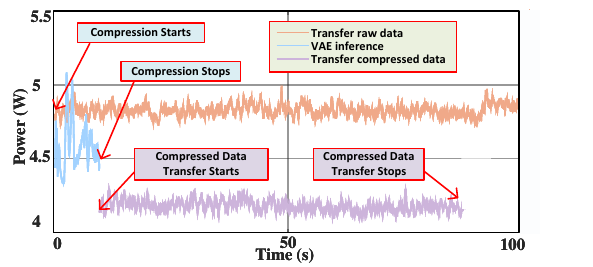}
\vspace{-20pt}
\caption{Power consumption graph for original signal transfer, compression runtime, and compressed signal transfer on Jetson Nano.}
\label{fig:power_jetson}
\vspace{-10pt}
\end{figure}

\subsection{Power Measurement}

To verify the power consumption reduction using the compression model compared to uncompressed data transfer, we conducted power measurements on Jetson Nano \cite{nvidiaJetsonNano} and Raspberry Pi 4 \cite{raspberrypiRaspberryModel}. Both devices are prevalent in IoT applications. We employed the Monsoon High Voltage Power Monitor (HVPM) \cite{msoonHighVoltage} to gauge real-time power usage while executing our code on compressed data sets and compared this to the power needed for transferring uncompressed data. 

The HVPM, capturing data at 5000Hz, powered and monitored the devices. For both the Raspberry Pi 4 Model B and Jetson Nano, the HVPM supplied power via the main channel voltage and ground, while monitoring through the USB ports. PowerTool, a proprietary software, was used alongside the Monsoon power monitoring system. The PowerTool software, with its capability to set start and stop checkpoints, ensured precise readings during inference script runtime. This precision eliminated unnecessary power fluctuations between data transfer and inference.

Post-measurement results revealed that transmitting compressed data consumed $\approx 27\%$ less power than raw data transfer, demonstrating our method's potential to cut energy consumption in resource-constrained IoT devices. Further compression refinements can potentially yield additional power savings. This study confirms our compression model's efficacy in reducing power consumption compared to conventional data transfer.

\textbf{Results}
When reviewing the overall impact of power draw on our edge devices we measured both original and compressed data workflows. For uncompressed data, we only measured simple transfer to the server via SCP. For compressed data, we combined power for VAE compression on-device and transfer of encoded data. This methodology provides a holistic view of the full compression pipeline's power requirements versus direct transmission.

\begin{figure}
\centering
\includegraphics[width=0.5\textwidth,trim = 40 30 120 30, clip]{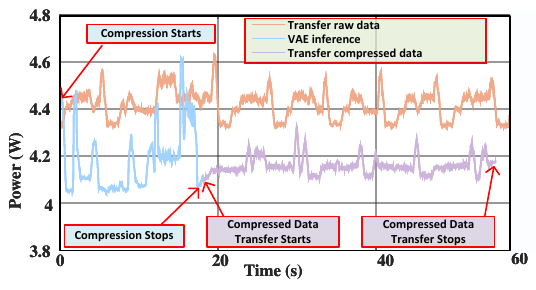}
\vspace{-25pt}
\caption{Power consumption graph for original signal transfer, compression runtime, and compressed signal transfer on Raspberry PI}
\label{fig:power_raspi}
\vspace{-15pt}
\end{figure}

As seen in Fig \ref{fig:power_jetson} \& \ref{fig:power_raspi}, there are 3 distinct power measurements shown in the graph. The combination of the blue measurement  (VAE inference) and the purple measurement, (compression data transfer) represents the total energy of the compression pipeline. The orange measurement (raw data transfer) represents the total energy of the raw signal transfer at the beginning of the pipeline without compression. From this data, we calculated the total energy expenditure in Joules based on the total time and power consumed.

\textbf{Jetson Nano} The Jetson Nano's total energy expenditure for the compression pipeline was 292.23J with 2.64W power expended, while the original transfer was 399.48 J with 3.95W power expended. In the compression pipeline energy breakdown, compression runtime was 9.67J and compression transfer was 282.56J. These results were calculated based on the average inference model runtime and transfer rate obtained by using the Jetson Nano's GPU capabilities. Overall our compressed model led to a 26.8\% reduction in energy expenditure and a 33.2\% reduction in power expenditure on the Jetson Nano's architecture. When using a generic USB 27,000 mAh battery pack with this device, our compression algorithm improved battery life from 34 hours and 10 minutes to 51 hours and 14 minutes. This is an increase in battery life of 17 hours and 4 minutes.

\textbf{Raspberry Pi 4 Model B} The Raspberry Pi's total energy expenditure for the compression pipeline was 209.60J with 3.66W power expended, while the original transfer was 243.92J with 4.00W power expended. In the compression pipeline energy breakdown, compression runtime was 68.89J and compression transfer was 140.71J. These results were calculated based on the average inference model runtime and transfer rate with the Raspberry Pi's CPU due to it not having a GPU like the Jetson Nano. Overall our compression model led to a 14.1\% reduction in energy expenditure and an 8.5\% reduction in power expenditure on the Raspberry Pi's architecture. When using a generic USB 27,000 mAh battery pack with this device, our compression algorithm improved battery life from 33 hours and 45 minutes to 36 hours and 53 minutes. This is an increase in battery life of 3 hours.

\vspace{-5pt}
\subsection{Medical Significance}
An exciting application of our work is in the field of deep brain stimulators, which are increasingly being to treat drug-refractory epilepsy (DRE) affecting a third of epileptic patients \cite{Pat1,Pat2}. In 2012, the FDA approved the first device, NeuroPace RNS, after clinical trials demonstrated both their ability to reduce seizure frequency and improve quality of life for DRE patients \cite{Pat3}. These devices work as closed-loop systems that algorithmically detect seizure activity based on EEG recordings via implanted electrodes and deliver electric stimulation to arrest seizure evolution and propagation \cite{Pat4}. Despite the increasing adoption of neural stimulators, their long-term use has been constrained by data storage and power management issues; wireless batteries must be recharged by patients every 8 hours, making it challenging to use these devices continuously \cite{Pat5}. Our approach to EEG signal compression has the potential of reducing this significant burden on patients while enabling faster transmission of data to the cloud where computations are increasingly being performed remotely, thereby reducing response latency. 

\section{Related Works}
\label{sec:relatedWorks}
It is crucial to monitor patients afflicted with conditions like epilepsy, as attacks can occur anywhere and at any time possibly causing major accidents and injuries. Furthermore, doctors benefit from having continuous health data collected throughout patients' daily activities beyond clinical visits to better understand their condition. 

Research into epileptic seizure predictions has yielded promising results. Harpale et al. utilized fuzzy inference and pattern-adapted wavelet transform (PAWT) \cite{10.1007/11578079_96} for EEG signal classification, achieving a 96.4\% accuracy with pre-seizure warnings 13-110 seconds before onsets \cite{HARPALE2021668}. Machine learning and deep neural networks have also achieved significant success \cite{10.1093/bib/bbw068, 6005992}. Usman et al. combined Empirical Mode Decomposition with neural networks, achieving early epilepsy prediction (32 minutes before onset) with 93\% sensitivity and 92.5\% specificity \cite{USMAN2021211}. Radwan et al. utilized CNN, LSTM, and GRUs for EEG-based personal identification, yielding 97.83\% accuracy \cite{10.1007/978-3-030-89701-7_3}. Mashid et al. proposed an entropy feature and ensemble learning-based seizure detection method, reaching 98\% accuracy for LS-SVM and Naive Bayes, and 94.5\% accuracy with KNN \cite{dastgoshadeh_detection_2023}. Rezaee et al. introduced an optimized seizure detection algorithm employing general tensor discriminant analysis for feature extraction and KNN classification, achieving 98\% accuracy \cite{rezaee_optimized_2016}. Shenglong Li et al. demonstrated gradient boosting's compatibility in medical data analysis in their work on orthopedic auxiliary classification \cite{10.1007_s00521-019-04378-4}. This progress was reinforced by Rasheed et al.'s review on the developments in ML-based algorithms for epileptic seizure prediction \cite{9139257}.

The challenge in current epileptic seizure prediction methods lies not in result accuracy but in the feasibility of prolonged sensing and the extraction of meaningful insights from compressed datasets. Compression models have been widely implemented across various medical domains to enhance classification performance. Vadori et al. introduced a compression technique for physiological quasi-periodic signals, achieving compression ratios of 35-, 70-, and 180-fold for PPG, ECG, and RESP signals respectively, thereby extending wearable device battery lifetimes \cite{vadori_biomedical_2016}. Gurkan et al. proposed a lossy compression method for EEG signals using a k-means clustering algorithm, attaining compression ratios as high as 93\% \cite{gurkan_eeg_2009}. Aghazadeh et al. developed a seizure detection algorithm for compressed EEG data on a multi-core architecture, achieving energy budgets of 18.4$\mu$J and 3.9$\mu$J for compression ratios of 24X, with accuracy reaching 90\%, utilizing non-linear SVM and dual linear SVM-based classification \cite{AGHAZADEH2020104004}. Aghababaei et al. detected epileptic seizures in compressed EEG data through ANOVA-selected features, employing decision trees, KNN, and discriminant analysis support machine methods, achieving 76.7\% accuracy at a 0.05 compression ratio \cite{AGHABABAEI2021114630}.
Compressing recorded signals opens a new area for enhancing mobile health systems.
Idrees et al. achieved superior data reduction while maintaining signal quality using a joint compression-prediction approach, enabling effective seizure prediction for wearables \cite{idrees2022edge}. Abdellatif et al. introduced a joint compression-prediction framework sustaining high seizure prediction accuracy even with notable compression ratios (e.g., only 0.6\% average loss for 1/2 to 1/16 compression) \cite{abdellatif2018automated}. Francesco et al. illustrated a use case employing an SVM classifier on low-power IoT devices for secure long-term monitoring and seizure detection \cite{7927716}. Navjodh et al. designed a Raspberry Pi-based device utilizing machine learning to identify traumatic brain injuries from a single-channel EEG signal, achieving over 90\% accuracy \cite{s21082779}. Shi Qiu et al. proposed a distributed edge computing-based seizure recognition algorithm design \cite{qiu_eeg_2022}.

In summary, improving long-term sensing and data recording methods is an active area of research that can pave the way for developing enhanced continuous health monitoring solutions for better healthcare management.
\section{Conclusion and Future Work}
\label{sec:futureWork}

In this work, we introduced a novel approach for compressing physiological signals, achieving an exceptional 1:293 ratio exceeding prior techniques. Critically, we verified efficacy by using the compressed data for 91\% seizure detection accuracy nearing 98\% from original signals. Our implementation demonstrated power savings on chips like Jetson Nano and Raspberry Pi versus uncompressed transfer. However, opportunities remain to further improve compression and match uncompressed performance in both accuracy and real-time inference. Though already advancing the state-of-the-art, continued enhancements could enable even greater effectiveness in analyzing physiological signals.

Some limitations still need addressing. For instance, our technique requires optimization for real-world EEG device constraints. Latent space representations may pose interpretability challenges versus visual review of raw EEGs. Also, our limited single-site study needs large multi-center trials to prove viability for mainstream clinical adoption. Meeting these challenges through ongoing research will facilitate the translation of our advanced compression approach into practical medical tools.

\bibliographystyle{plain}
\bibliography{reference}

\end{document}